\documentclass[oneside]{IEEEtran}
\usepackage[cmex10]{amsmath}
\usepackage{amssymb}
\usepackage{cite}
\usepackage{amsmath}
\usepackage{amssymb}
\usepackage{booktabs}
\usepackage{balance}
\usepackage{subfigure}
\usepackage{multicol}
\usepackage{graphicx}
\usepackage{epsfig}
\usepackage{epstopdf}
\usepackage{algorithm}
\usepackage[colorlinks,linkcolor=red]{hyperref}
\usepackage{algorithmic}
\usepackage{bm}
\usepackage{amsmath, amssymb, amsthm}
\usepackage{array}
\usepackage{makecell} 
\usepackage{slashbox}
\usepackage{multirow}
\usepackage{float}
\usepackage{multicol,multicap,graphicx}
\usepackage{1-in-2}
\usepackage{mathrsfs}
\usepackage{url}
\usepackage{ifpdf}
\usepackage[bf]{caption2}
\usepackage{color}
\usepackage{1-in-2}
\renewcommand\tablename{Table}
\renewcommand\thetable{\arabic{table}}

\renewcommand{\figurename}{Figure}

\ifCLASSINFOpdf
\else
\fi

\begin{document}

\title{Physiological Signal-Driven QoE Optimization for Wireless Virtual Reality Transmission}
\author{Chang Wu$^{\dag}$, Yuang Chen$^{\dag}$, Yiyuan Chen, Fengqian Guo$^{\ast}$, Xiaowei Qin, and Hancheng Lu,
}
\maketitle

\section*{Abstract}
Abrupt resolution changes in virtual reality (VR) streaming can significantly impair the quality-of-experience (QoE) of users, particularly during transitions from high to low resolutions. Existing QoE models and transmission schemes inadequately address the perceptual impact of these shifts. To bridge this gap, this article proposes, for the first time, an innovative physiological signal-driven QoE modeling and optimization framework that fully leverages users' electroencephalogram (EEG), electrocardiogram (ECG), and skin activity signals. This framework precisely captures the temporal dynamics of physiological responses and resolution changes in VR streaming, enabling accurate quantification of resolution upgrades' benefits and downgrades' impacts. Integrated the proposed QoE framework into the radio access network (RAN) via a deep reinforcement learning (DRL) framework, adaptive transmission strategies have been implemented to allocate radio resources dynamically, which mitigates short-term channel fluctuations and adjusts frame resolution in response to channel variations caused by user mobility. By prioritizing long-term resolution while minimizing abrupt transitions, the proposed solution achieves an 88.7\% improvement in resolution and an 81.0\% reduction in handover over the baseline. Experimental results demonstrate the effectiveness of this physiological signal-driven strategy, underscoring the promise of edge AI in immersive media services.

\IEEEpeerreviewmaketitle

\mathfootnote{\centering $^{\dag}$ These authors are co-first authors. $^{\ast}$Corresponding author. \footnotesize{Chang~Wu, Yuang~Chen, Yiyuan~Chen, Fengqian~Guo, Xiaowei~Qin, and Hancheng~Lu are with University of Science and Technology of China. Hancheng Lu is also with Deep Space Exploration Laboratory, Hefei 230088, China.}}

\section{Introduction}
\IEEEPARstart{V}{irtual} reality (VR) has emerged as a transformative medium for immersive digital experiences, driven by its capacity to deliver high-resolution 360$^{\circ}$ videos with ultra-low motion-to-photon (MTP) latency \cite{yaqoob2020survey, 10907861}. While this technology enables unprecedented engagement in applications ranging from event viewing to interactive education, its reliance on wireless transmission poses critical challenges. The uncompressed data rates exceeding 1 \emph{Gbps} and latency thresholds below 20 \emph{ms} impose stringent demands on network infrastructure, particularly in mobile scenarios where channel fluctuations and user mobility degrade service consistency. Traditional quality of service (QoS) metrics (e.g., bandwidth, jitter, and packet loss) provide necessary but insufficient insights into user satisfaction, necessitating perceptual quality of experience (QoE) frameworks tailored to user's unique requirements \cite{anwar2020measuring,fei2019qoe,zuo2022adaptive}.

\par Existing studies have almost failed to consider the impact of asymmetric perception of resolution conversion on QoE. For example, current adaptive bitrate (ABR) algorithms, which are effective for conventional video services, exhibit inherent limitations when applied to VR environments. The ABR schemes based on network and playback state perception provide the highest possible video quality without interruption through reactive bitrate adjustments \cite{johansson2014self,maura2024experimenting}. Some objective studies have shown that even switching from high to medium resolution can lead to significant degradation in experience compared to continuous medium-quality playback \cite{bampis2021towards}. This phenomenon aligns with principles of behavioral psychology, where negative experiences disproportionately influence overall satisfaction compared to positive counterparts. However, existing QoE models generally quantify quality through metrics such as Mean Opinion Score (MOS), completely ignoring the inherent principles of how the direction and degree of resolution changes affect video quality \cite{agarwal2022qoe}.

\par The architectural separation between network provisioning and application-layer adaptation further exacerbates these challenges. The radio access network (RAN) schedulers in most conventional systems allocate resources based solely on physical-layer channel state information (CSI), disregarding the perceptual urgency of video frames \cite{3gpp_technical_38214}. Concurrently, ABR controllers at the transport layer make suboptimal bitrate decisions using delayed network feedback, creating a latency-induced mismatch between encoding rates and available channel capacity \cite{bentaleb2018survey}. This disjointed optimization results in oscillatory quality adjustments, resource underutilization, and excessive resolution switches. The issues are amplified by VR's extreme throughput requirements and human sensitivity to visual inconsistencies. To address these challenges, it is imperative to develop effective adaptive streaming solutions that ensure proper RAN capacity provisioning and source resolution adaptation while incorporating an evaluation model that reflects human sensory experience. By providing suitable RAN capacity, overall capacity efficiency is enhanced and the complexity of source-end adaptation is reduced \cite{agarwal2022qoe,yeznabad2024qoe}. Through improved responsiveness in video resolution adaptation, the system can more accurately match encoding bitrates to network conditions, minimizing both capacity waste and under-provisioning. Integrating VR evaluation models that consider perceptual factors, like user sensitivity to resolution changes, leads to more user-centric strategies. This combined approach allows for necessary dynamic resolution adjustments that respond swiftly to network variations while minimizing the negative perceptual impact of quality switching, ultimately enhancing the VR experience.

\par In this research context, this article proposes for the first time an innovative physiological signal-driven QoE model to drive the optimization of VR transmission. In particular, we analyze the intensity of physiological changes corresponding to different resolution switches by monitoring users' physiological indicators, including electroencephalography (EEG), electrocardiography (ECG), and electrodermal activity (EDA) when users watch VR content with different resolutions. This endows the proposed QoE model with the ability to quantify the perceived impact of video quality upgrades and downgrades, and assign appropriate penalties for resolution changes for different directions and levels. Integrating the proposed QoE model into a deep reinforcement learning (DRL) framework enables RAN to provide appropriate channel capacity for frame transmission and helps adjust frame resolution on time. On the one hand, radio air-interface resources are intelligently allocated to match the rate and urgency of upper-layer frame transmission \cite{agarwal2022qoe}. On the other hand, RAN determines adjustment factors to determine the resolution of subsequent frames. By considering both network conditions and physiological responses, our framework strikes a balance between efficient transmission and optimal user experience, minimizing abrupt quality switches and enhancing immersion.

\par The remainder of this article is organized as follows. The system architecture and key challenges of wireless VR video transmission are first described. Then, we introduce a physiology-driven VR experiment and present a DRL-based joint optimization scheme, followed by performance evaluation and analysis. Finally, we conclude this article and discuss future issues.

\begin{figure}[!t]
    \centering
    \includegraphics[width=3.45in]{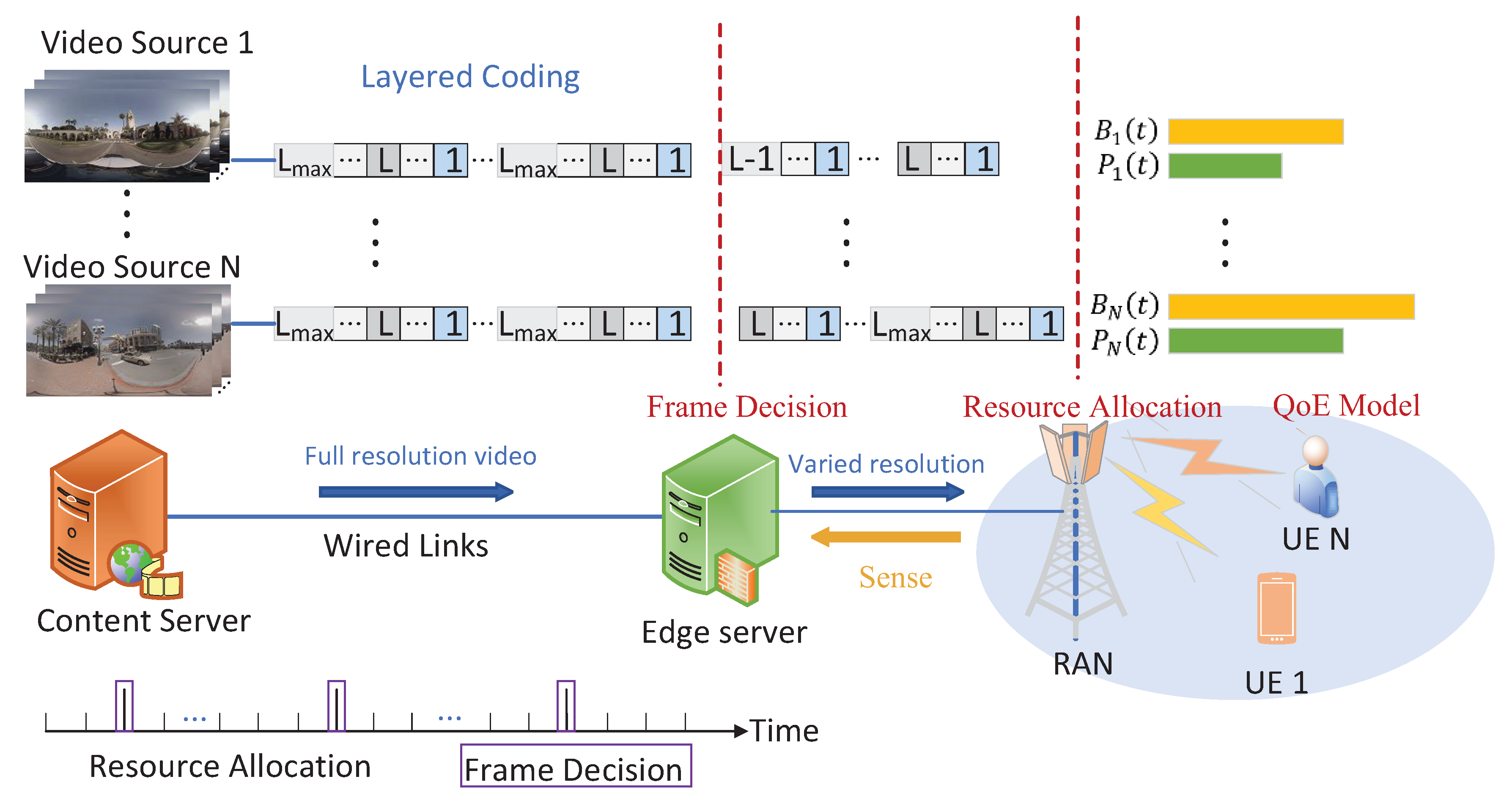}
    \caption{The structure and work process of the wireless VR system.}
    \label{fig:strucure_and_work_process}
    \vspace{-1em}
\end{figure}

\section{System Architecture and Key Technologies}

\par As illustrated in Figure \ref{fig:strucure_and_work_process}, we present the structure and workflow of the wireless VR video transmission system. This article focuses on downlink VR streaming scenarios, with a particular emphasis on the wireless transmission stage between the edge RAN and users. The fundamental challenge arises from the discrepancy between high-capacity wired links in the core network and comparatively constrained wireless channels at the edge. Consequently, the layered-coded VR frames that may contain the highest resolution are queued at the RAN before being delivered to each user over the air interface. The effective resolution received depends on the RAN's assessment of prevailing channel conditions or user preferences within channel capacity limits, which introduces variability in user-perceived QoE since different resolutions trigger varying levels of immersion and satisfaction. To efficiently utilize the limited wireless resources while consistently streaming high-quality VR content, edge intelligence must be employed in the RAN. Specifically, it is imperative to develop advanced decision-making mechanisms that harmonize resource allocation, frame resolution adaptation, and user-centric QoE. Moreover, device mobility and heterogeneous demands necessitate dynamic adaptation, underscoring the importance of resource allocation for stringent requirements of VR streaming. Below, we discuss three key challenges essential to implementing the robust wireless VR streaming system: the designed QoE model, frame resolution selection, and radio resource allocation.

\par \textbf{QoE Model:} An accurate QoE model is the cornerstone of wireless VR streaming since it transforms complex performance metrics into meaningful insights about user satisfaction. Unlike conventional video streaming, VR streaming significantly amplifies the impact of visual and interactive disruptions \cite{anwar2020measuring,fei2019qoe}. Metrics such as perceived realism, MTP latency, and smoothness become critical in shaping the VR experience. However, existing QoE metrics usually overlook users' subjective responses to dynamic resolution changes, where abrupt downgrades can disrupt immersion and abrupt upgrades may momentarily distract from the current view. Therefore, an accurate QoE model for VR streaming must account for these perceptual factors and be able to provide finer-grained reflections to reveal how each resolution transition affects user comfort and engagement.

\par \textbf{Frame Resolution Decision:} Frame resolution decisions are the core of immersive VR experiences. The system must carefully determine when to switch resolutions and how to select the appropriate layer from the queued video frames at the RAN. 
Abrupt transitions can disrupt the user's sense of presence, while overly conservative strategies may deliver suboptimal clarity \cite{anwar2020measuring, maura2024experimenting}. This balancing act is complicated by the fact that VR users look around in a 360$^{\circ}$ environment, causing the region of interest (RoI) to change over time. Consequently, the frame resolution decision process must incorporate real-time feedback, channel estimation, and predictive algorithms to optimize user's experience under strict resource constraints.

\par \textbf{Radio Resource Allocation:} The RAN must flexibly schedule limited spectrum, power, and time resources among users and frames, as users and frames have varied QoE and urgency requirements. Inadequate allocation can create bottlenecks, leading to frame loss or degraded resolution that significantly diminishes the user's immersion. In addition to traditional metrics like throughput and fairness, VR demands awareness of frame deadlines, quality sensitivity, and user mobility. Moreover, resource scheduling decisions directly influence how frame resolution adaptations unfold, highlighting the interdependence between RAN provisioning and QoE-driven encoding strategies \cite{agarwal2022qoe,yeznabad2024qoe}. Overcoming these interlinked challenges requires joint optimization approaches that maximize system efficiency without sacrificing user experience.

\begin{figure}[!t]
    \centering
    \includegraphics[width=3.45in]{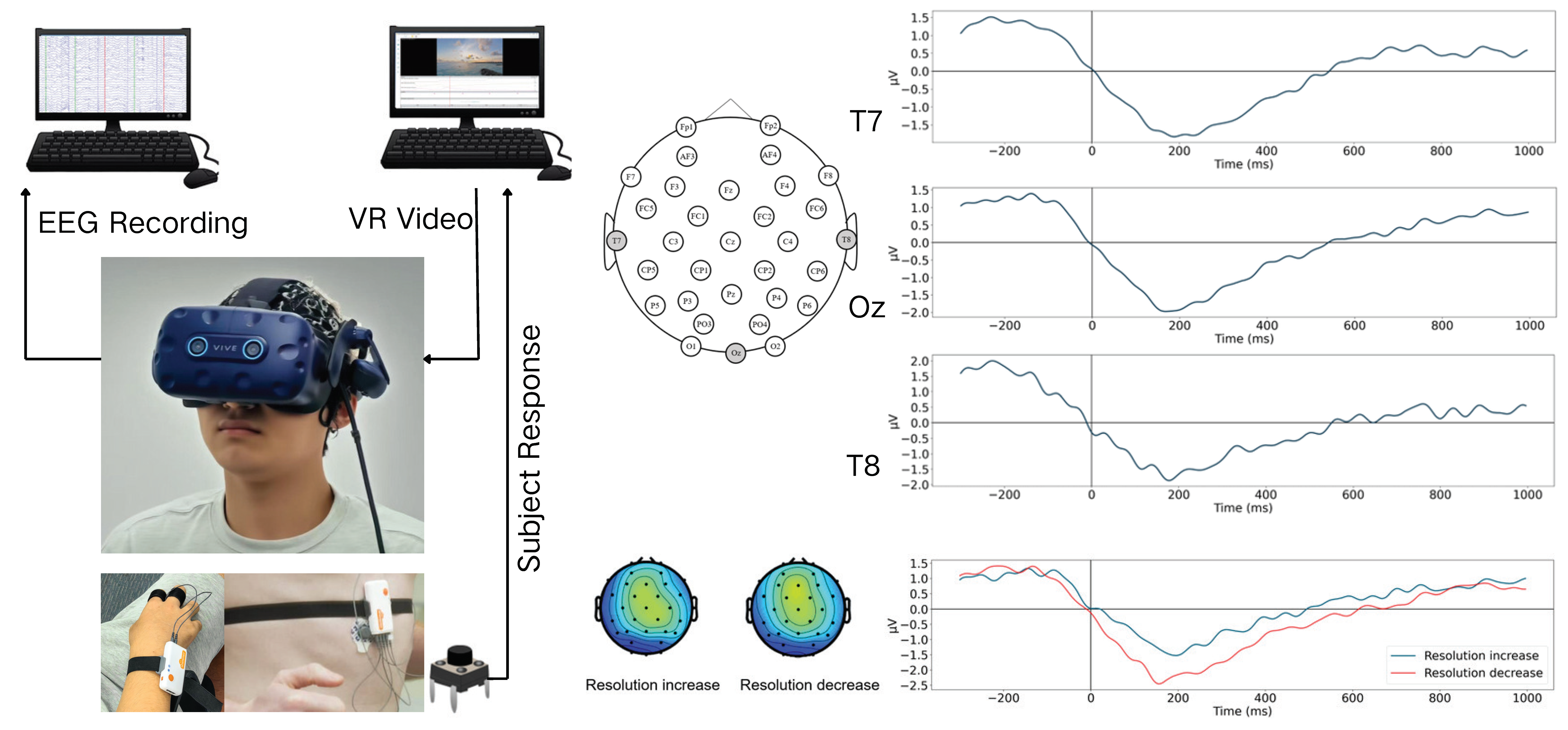}
    \caption{Framework of VR physiological experiment system and related results.}
    \label{fig:VR_PHY_EXP}
    \vspace{-1em}
\end{figure}

\section{Physiological Signal Analysis and QoE Model Construction}

\par For the first time, this article establishes a quantitative relationship between resolution variations and physiological responses by analyzing users' bioelectrical signals, quantifying the impact of VR streaming's abrupt resolution changes on QoE. A total of 19 healthy participants (11 males, 8 females, average age 23) were recruited. The experiment utilized \emph{HTC VIVE PRO EYE} to display 29 carefully selected natural-scene VR videos (40 seconds each, original resolution 8K). During playback, resolution levels (8K, 4K, 1080P, 720P, 480P) were randomly switched every 8 seconds, with each participant exposed to 116 resolution-change stimuli. A 10-second black screen was displayed before each video to alleviate visual fatigue. Participants pressed a feedback button upon perceiving resolution changes to tag valid stimuli.

\par As shown in Figure \ref{fig:VR_PHY_EXP}, physiological data were collected synchronously using a 32-channel \emph{Neuracle Neusen W} wireless EEG system and \emph{Shimmer 3} sensors (for galvanic skin response (GSR) signals)\footnote{These raw data can be obtained via \href{https://rec.ustc.edu.cn/share/fb4a56a0-650d-11f0-ac29-33d1eecb5cd8}{LINK-1} by password \texttt{27mh}.}. EEG signals were preprocessed with a 1-20 Hz bandpass filter, downsampled to 250 Hz, and artifacts (e.g., ocular, muscular) were removed via independent component analysis. The GSR signals were interpolated using bilinear methods, with trials retained only if signal-to-noise ratios exceeded 20 dB. Results revealed that resolution changes elicited a prominent N200 event-related potential (ERP) in EEG signals, peaking at 200 \emph{ms} post-stimulus over temporal (T7/T8) and occipital (Oz) regions, reflecting visual processing and attentional allocation. Notably, N200 amplitudes were significantly larger during resolution degradation (high-to-low transitions) compared to enhancement (low-to-high transitions) ($p < 0.001$), indicating heightened sensitivity to quality deterioration. To evaluate the effect of resolution transition magnitude, changes were categorized into large ($>$2 levels) and small ($\leq$ 2 levels) jumps. Both resolution enhancement and degradation conditions exhibited stronger N200 responses for larger jumps. Scalp topographic maps demonstrated concentrated occipital energy during large jumps versus diffuse distributions for small jumps, suggesting intensified visual processing for significant quality shifts. Statistical validation confirmed the significance of jump magnitude on N200 amplitudes (enhancement: $p < 0.001$; degradation: $p < 0.05$). A linear discriminant analysis (LDA) classifier was constructed using time-windowed EEG features from the Oz channel (1-second post-stimulus) \cite{vidaurre2009time,blankertz2011single}. The Signed-$r^2$ method identified discriminative temporal intervals, achieving an average area under curve (AUC) of 74.12\% across 19 participants, confirming neural specificity to resolution-change directions. For GSR analysis, a dual-path model integrating phasic (transient) and tonic (baseline) features attained an average AUC of 78.10\% across 16 participants. Despite environmental susceptibility, GSR signals effectively captured autonomic responses to resolution shifts.

\par \textbf{The real-world multimodal physiological signals challenge the asymmetry assumption in traditional QoE models}, highlighting the necessity of incorporating directional and magnitude-dependent weighting. In particular, physiological responses triggered by solution degradation are stronger than those caused by enhancement, and larger jumps amplify neural and electrodermal activities. Therefore, QoE models should prioritize physiological metrics and place particular emphasis on sensitivity to resolution degradation events and large-scale transitions to refine the evaluation of adaptive resolution strategies in VR environments.

\vspace{-1em}

\section{Joint Capacity Provision and Resolution Decision Optimization}

\par This article proposes an edge intelligence decision-making framework, which considers specific QoE models to achieve the highest possible QoE performance with limited resources through appropriate capacity provision and resolution adaptation. The system model consists of a BS equipped with edge computing capabilities, streaming live VR services to $N$ users within its coverage area, as shown in Fig. \ref{fig:strucure_and_work_process}. Layered VR video data based on scalable coding is cached on the RAN side and transmitted to end users via the air interface. To focus on RAN optimization, four assumptions are introduced: 1) All users receive content at identical frame rates; 2) Orthogonal frequency division multiplexing (OFDM) eliminates inter-user interference; 3) The server-to-5GC link uses high-capacity fiber optics with fixed latency; 4) Video frame packets have uniform sizes. Through minor parameter adjustments, the adaptability of the system model remains adaptable to generalized scenarios.

\par The primary challenge lies in balancing limited wireless resources to deliver stable, high-quality VR services while maintaining fairness among users. To this end, this article proposes a novel dual-timescale control architecture, which decouples fast-paced radio resource management from slower, QoE-aware resolution adaptation. In particular, based on the buffer state, packet deadline, and channel conditions observed by the edge server, the resource allocation is performed at the millisecond timescale for each time slot to determine capacity provisioning. Bandwidth is allocated dynamically among users, with priority given to packets approaching deadlines to mitigate frame failures. This granular control ensures the timely delivery of queued frames while adhering to fairness constraints. In addition, the resolution adaptation module evaluates aggregated network performance metrics, including buffer occupancy and historical channel capacity at the frame-level timescale (tens of milliseconds). Drawing on these insights, the system dynamically adjusts the resolution of upcoming video frames. Resolution decisions are piggybacked onto acknowledgment packets returned from users, establishing a closed-loop control mechanism that aligns encoding strategies with observed network behavior.

\par To address the intertwined temporal dependencies and high-dimensional action space inherent in joint resource allocation and resolution control. DRL emerges as a natural choice, enabling the system to learn optimal control policies through continuous interaction with the network environment. This article employs two collaborative DRL agents, where the scheduling \& utility (SU) agent performs bandwidth allocation per timeslot to maximize frame delivery success under fairness constraints. In contrast, the resolution scaling (RS) agent executes frame-level resolution adjustments based on predicted network conditions to optimize perceptual QoE. State representation integrates real-time network observation (NO) with video transmission status (VTS). The state of the RS agent is frame-granular, with its NO including CSI and buffer occupancy, and its VTS including delivery success metrics, historical resolution, frame delay, and packet loss rate. The state of the SU agent is based on time slots, with its NO augmented by instantaneous transmission volume and historical action, and its VTS including the remaining time of the frame. Action spaces reflect agent-specific responsibilities. The SU agent outputs a continuous resource allocation ratio among users, while the RS agent selects a discrete resolution switch from several predefined options. The reward design strikes a balance between transmission efficiency and perceived quality. The reward for the SU agent is calculated based on frame delivery success, resource efficiency, and the rate required for the remaining data of the frame. The reward of the RS agent includes resolution level reward, failure penalty, and frame resolution degradation penalty, which is proportional to the level. Moreover, the proposed physiological signal-driven QoE feedback quantifies user discomfort during resolution transitions, which directly guides reward calculation.

\par The proximal policy optimization (PPO) algorithm is adopted as the core optimization framework due to its demonstrated advantages in stability and sample efficiency over other DRL architectures (e.g., deep Q-networks or actor-critic variants \cite{schulman2017proximal}). The clipped objective function incorporated by PPO effectively constrains policy updates within a trust region and ensures stable convergence characteristics. This is crucial for wireless network environment plagued by CSI fluctuations and heterogeneous VR video resolution demands, which together form a non-stationary learning scenario \footnote{The code of our article is available at \href{https://github.com/ChangWu98/LiveVR\_PPO}{LINK-2}}.


\begin{figure}[!t]
    \centering
    \includegraphics[width=3.45in]{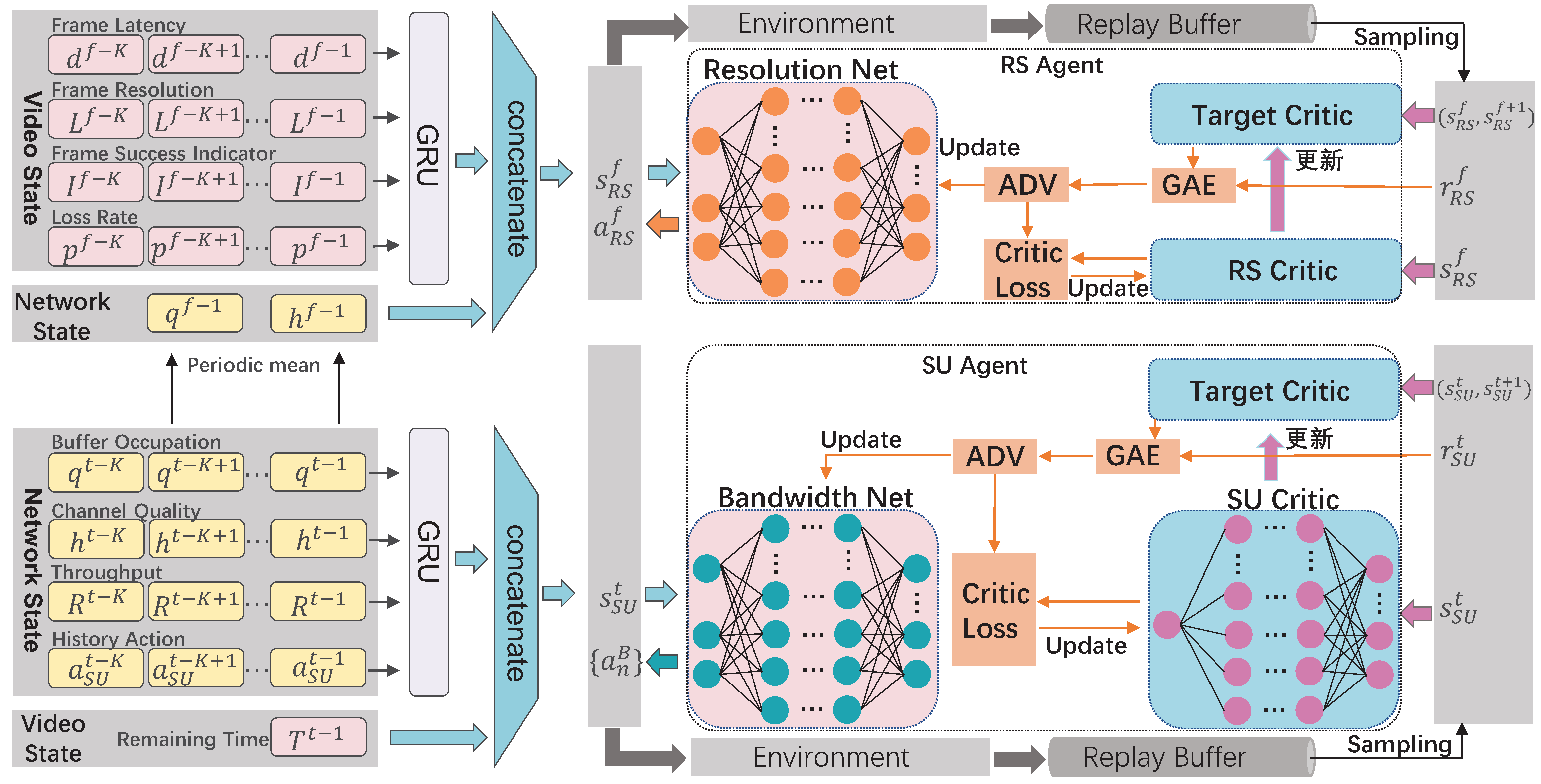}
    \caption{Multi-agent DRL framework with GRU as feature extraction enhancement.}
    \label{fig:DRL_Structure}
    \vspace{-1.5em}
\end{figure}

\par As illustrated in Figure \ref{fig:DRL_Structure}, a dedicated network architecture is created for each agent in our DRL framework. The feature extraction part module remains universal, in which two parallel gated recurrent unit (GRU) layers process temporal sequence data from the application layer of the RS agent and the network layer state of the SU agent, respectively. Since these data directly reflect state transitions and determine the behavioral rewards, the extraction of sequence features by GRU components enhances the perception of state transitions, improving decision-making performance. The GRU outputs, concatenated with the remaining tolerable frame delay, are fed into a dense layer with 64 neurons and \texttt{tanh} activation in the decision network to capture cross-domain dependencies. The RS agent employs \texttt{softmax} activation for selection across seven resolution levels, while the SU agent utilizes \texttt{sigmoid} activation to generate normalized resource allocation ratios among users. The critic network mirrors this architecture but replaces the final activation with a linear output for value estimation. To accelerate convergence, we implement prioritized experience replay and a dynamic learning rate scheduler adapted to training progress. During training, the SU agent is first trained under diverse channel conditions and frame sizes to attain capabilities in resource-efficient scheduling and frame-urgent awareness. Then, the RS agent is trained based on per-frame decisions and interactions, with the trained SU agent providing the optimal resource scheduling-induced capacity provisioning per time slot. The resulting network states and user QoE feedback are stored for policy updates. This co-design enables collaborative adaptation. Resolution selection considers predicted resource availability, while resource scheduling decisions anticipate forthcoming capacity demand. By unifying these independent control loops, the proposed DRL framework significantly reduces the oscillatory behavior of both capacity and frame resolution.

\par The integration of physiological signal-driven QoE metrics introduces human-centric safeguards, where resolution downgrades are only permitted when absolutely necessary, and significant quality degradation triggers disproportionately high penalties. Conversely, during network recovery phases, gradual upgrades are encouraged by aligning technical optimization with perceptual comfort. The proposed DRL architecture demonstrates how edge intelligence bridges the gap between network-centric metrics and human perception, paving the way for truly immersive wireless VR experiences. 

\section{Performance Evaluation}

\par In this section, the proposed dual-timescale joint optimization scheme is validated through extensive comparative simulations with the current state-of-the-art schemes. Firstly, wireless channel time-series data is measured based on the established wireless channel model and user mobility model. This article developed a wireless channel simulation module based on the \texttt{3GPP} guidelines on VR performance measurement, where multiple users randomly walk within a $100 m \times 100 m$ square area centered on the BS. For this module, the noise power is assumed to be -174 dBm/Hz, the carrier frequency is 3.5 GHz, and the antenna heights at the base station and users are $4 m$ and $1.5 m$, respectively. The standard deviation of shadow fading is assumed to be 8 dB with a shadowing correlation distance of $50 m$. In video data transmission optimization, the frame rate is uniformly set to 50 FPS. Thus, the maximum frame transmission delay can be specified from the frame interval down to smaller durations. Frame sizes are obtained from encoding several video segments using \texttt{ffmpeg}. For the proposed DRL architecture, both the actor and critic networks use two fully-connected hidden layers, each containing 64 neurons. The neural network parameters are initialized orthogonally, and the generalized advantage estimator (GAE) method is used to compute the policy gradient. The learning rates for the actor and critic networks are initialized to $3 \times 10^{-4}$. A linear learning rate decay schedule is used to linearly reduce the learning rate from the initial value to zero as the training steps increase. The discount factor for the SU agent is set to 0.99, while that for the RS agent is set to 0.9, due to the stronger correlation between adjacent slots. Based on training performance, the GAE parameters for the SU and RS agents are set to 0.95 and 0.8, respectively. The replay buffer size and batch size are set to 2048 and 64, respectively. Moreover, the simulations are implemented using Python 3.9 and PyTorch 2.6 on an Apple M1 CPU running at 3.2 GHz and 16 GB RAM.

\begin{figure}[!t]
    \centering
    \includegraphics[width=3.45in]{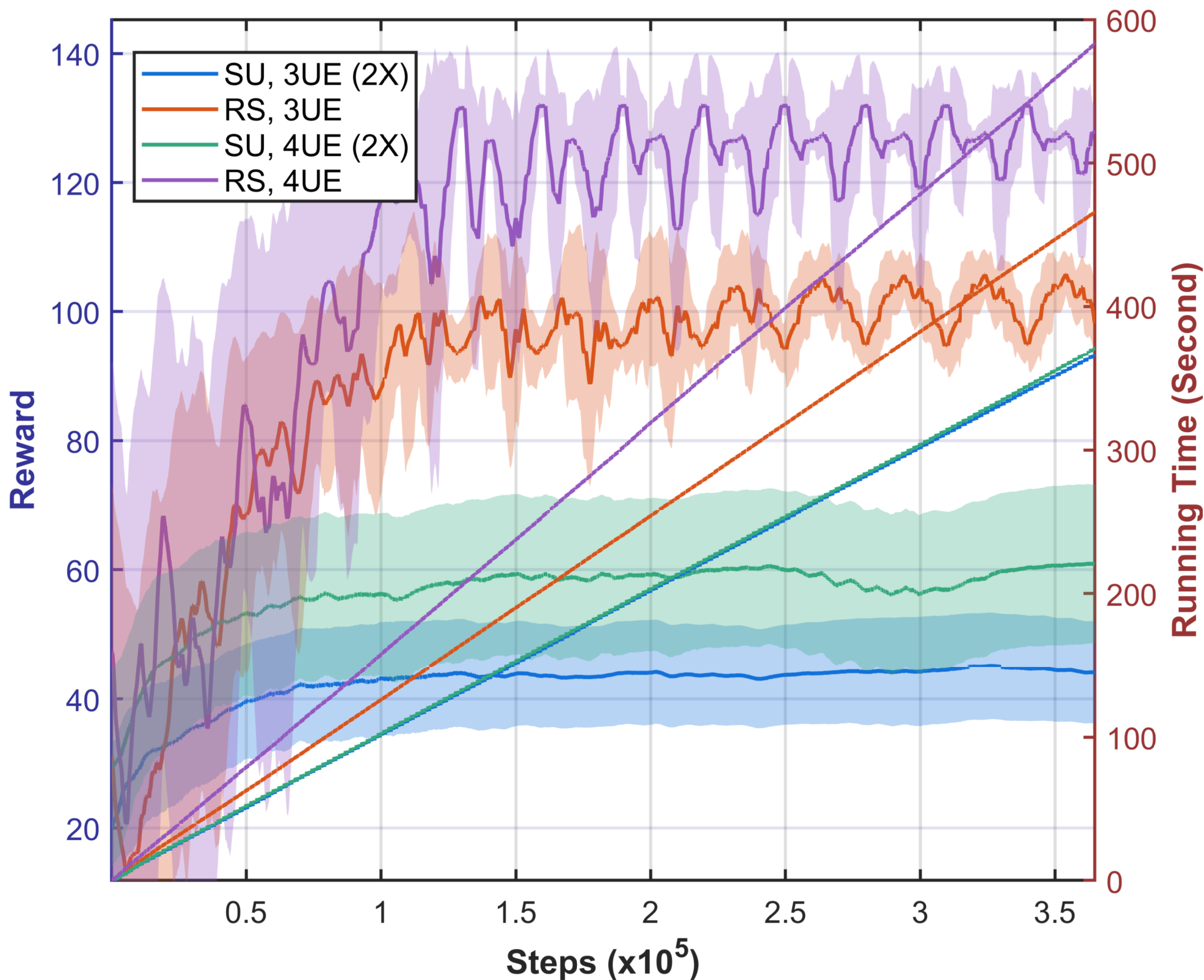}
    \caption{Training curves and running time of the proposed scheme for different user numbers.}
    \label{fig:DRL_Training}
    \vspace{-1.5em}
\end{figure}

\par
\begin{figure}[!t]
    \centering
    \includegraphics[width=3.45in]{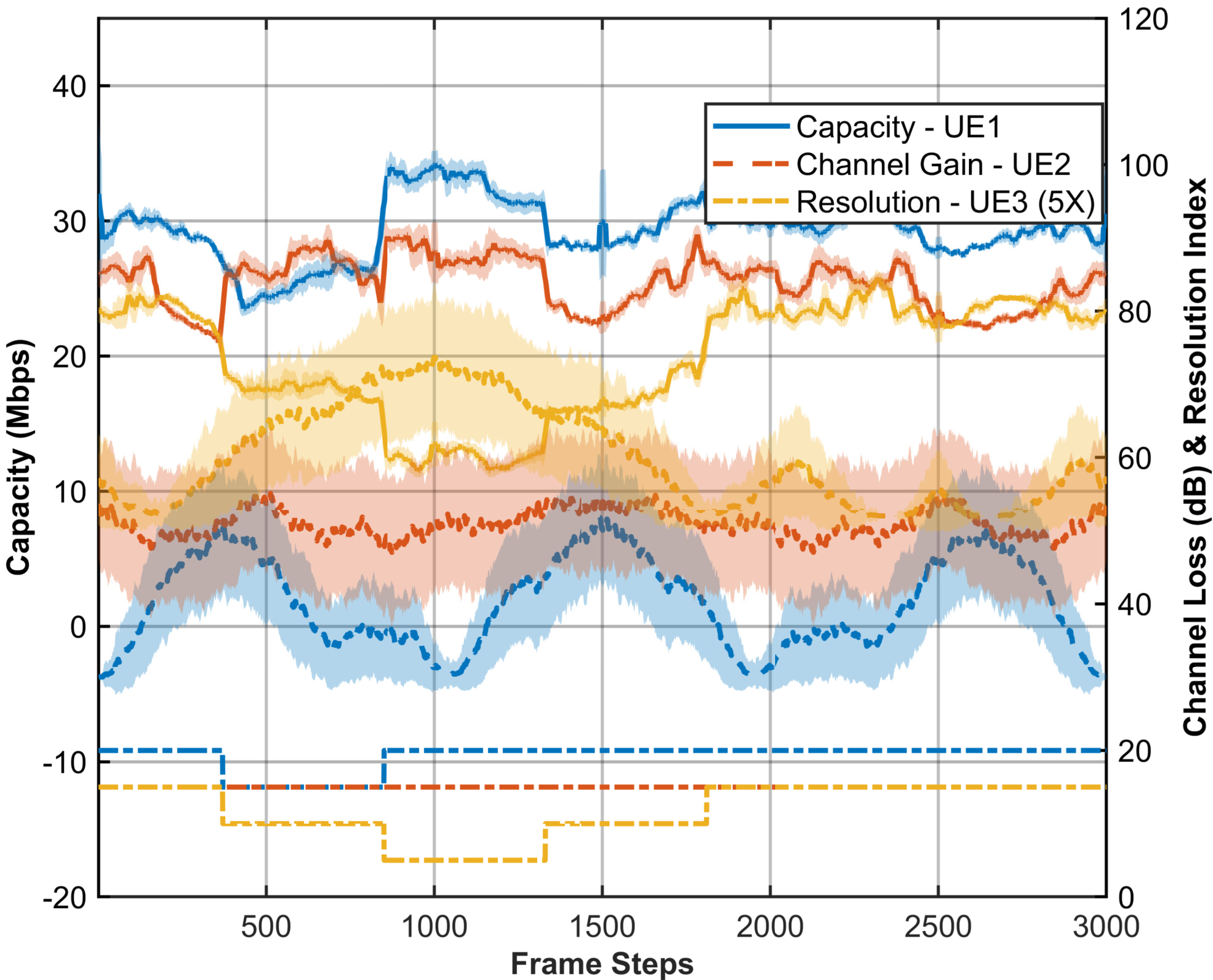}
    \caption{The effect of joint decision-making between two agents in time-varying channels.}
    \label{fig:time_seq}
    \vspace{-1.5em}
\end{figure}

 We first evaluate the convergence and cost of the proposed solution, including the number of iterations and runtime required for convergence. As shown in Figure \ref{fig:DRL_Training}, an increase in the number of system users slightly increases the training time and convergence duration, owing to the larger data space and increased parameter count. Furthermore, a larger user scale necessitates more interactions between the agent and the environment to explore favorable directions. It should be noted that the periodic fluctuation in the reward curve of the RS agent stems from the cyclic reuse of the channel. The mean reward of the SU agent is multiplied by a factor of 2 for enhanced visualization. The collaborative decision-making effect of the trained RS and SU agents is shown in Figure \ref{fig:time_seq}. Three variables are plotted on a dual-y-axis diagram for clear observation. The left y-axis represents the capacity data for three users, resulting from the SU agent's decisions, while the right y-axis simultaneously depicts the total wireless channel loss for the users and the resolution level decided by the RS agent under varying channel conditions. Note that the resolution level is multiplied by a factor of 5 for clear observation.

\par On the one hand, the user with consistently better overall channel conditions maintains a higher video resolution, which aligns with the objectives of resource efficiency and overall system optimality. On the other hand, users with similar channel conditions are always allocated the same video resolutions, adhering to the principle of fairness. Most significantly, the resolution level remains stable despite highly fluctuating channel conditions around the mean value across time slots. It changes only when the mean channel quality variation exceeds a certain threshold, achieving resolution adaptation. This meaningful effect stems from the RS agent's grasp of the relationship between resolution level and overall channel quality, coupled with the SU agent providing stable and efficient capacity supply across frames. During performance evaluation, the frame success rate exceeds 99\%, effectively validating the practical significance of the dual-timescale agents' joint decision-making. To validate the effectiveness of the proposed PPO-based DRL optimization scheme, three baseline schemes are considered.

\textbf{SU Agent Only:} Resource allocation employs either equal division or proportional-fair (PF) schemes, which verifies the effectiveness of the SU agent considering frame tightness in ensuring frame success.

\textbf{Congestion Control (CC) Only:} Taking Ericsson's SCReAM as an example \cite{johansson2014self}, this contrasts the performance of the RS agent in terms of frame stability.

\textbf{Different Frame Resolution Change Penalty Weights:} This verifies the performance of the RS agent in balancing video frame quality and its continuity.

\begin{figure}[!t]
    \centering
    \includegraphics[width=3.4in]{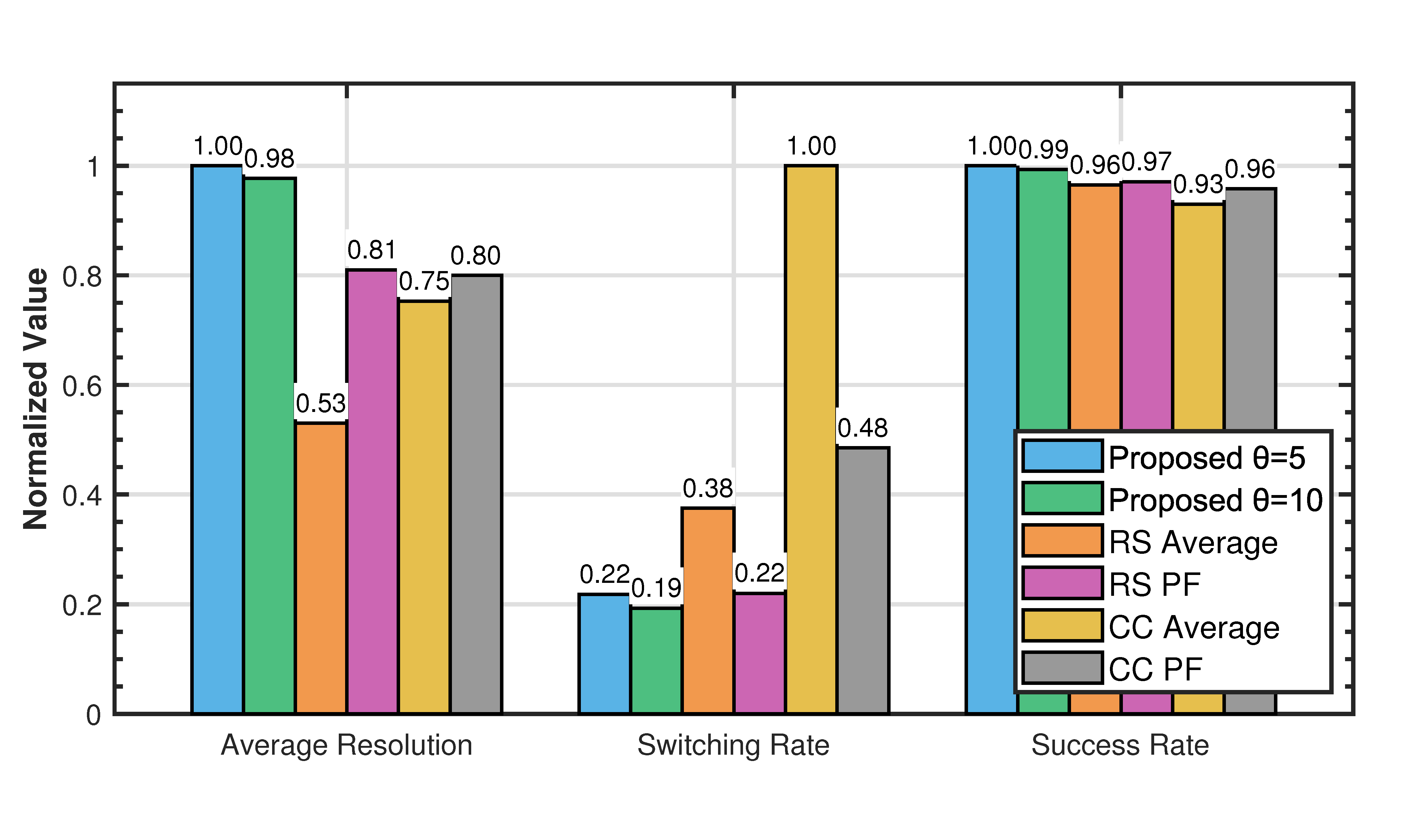}
    \caption{Performance comparison of different schemes in terms of average video frame resolution, switching rate and successful transmission probability.}
    \label{fig:performance_comparison}
    \vspace{-1.5em}
\end{figure}

\par As shown in Figure \ref{fig:performance_comparison}, the performance of our proposed scheme and the comparative schemes is contrasted in terms of average video quality, resolution switching, and frame success rate, where $\theta$ denotes the penalty value for resolution changes. All performance results are normalized for convenient comparison. It can be observed that our proposed dual-timescale joint optimization framework achieves the best performance in all three aspects. Compared to the average resource allocation scheme, the average video resolution is improved by up to 88.7\%, and the resolution switching rate is reduced by up to 81.0\%, validating the superior performance of the proposed scheme in delivering stable and high-quality video streams. Conversely, the decoupled upper-layer congestion control with an underlying equal resource allocation scheme achieves nearly the worst performance, due to the high fluctuation in wireless capacity caused by equal resource division and the hysteresis of upper-layer congestion control regulation. Further analysis reveals that when a larger penalty is imposed on resolution switching (larger $\theta$), the scheme attains the best performance in switching rate but poorer performance in the other two aspects compared to smaller $\theta$. This occurs because the emphasis on maintaining the current resolution level leads to infrequent switching, which easily causes frame failures due to exceeding the capacity supply. Moreover, owing to our high priority on frame success, the system tends to make more conservative decisions on frame resolution levels, resulting in the RS agent with equal resource allocation exhibiting the worst performance in average resolution. Additionally, PF-based resource allocation balances resource efficiency and inter-user fairness, and leads to smoother transmission capacity compared to equal resource allocation. As a result, our scheme outperforms the corresponding equal allocation scheme in all three aspects.

\section{Conclusion and Future Work}

\par The pursuit of seamless wireless VR experience is fundamentally challenged by the fragility of human visual perception and the volatility of radio channels. Although traditional adaptive streaming media focuses on network-level metrics, our work suggests that due to innate neurophysiological sensitivity, even if technically necessary, abrupt resolution degradation disproportionately reduces QoE. In this article, we pioneered an innovative physiological signal-driven QoE framework that effectively addresses this challenge by quantifying the asymmetric effects of resolution conversion through multimodal physiological signal analysis, revealing for the first time the level of user aversion to downgrades versus upgrades. By integrating these insights into an edge based dual agent DRL architecture, our solution achieves joint optimization of millisecond-scale radio resource allocation and frame-scale resolution adaptation. This collaborative design bridges the traditional gap between network provisioning and application-layer adaptation, improving average resolution level and resolution switching performance by 88.7\% and 80.7\%, respectively. This work validates that edge intelligence guided by human physiology can transform the rigid ``quality first'' stream into a perceptual resonance experience, establishing a blueprint for human-centric immersive networks. As an emergent paradigm, there remains many open issues for further study in the future.

\textbf{More Comprehensive QoE Models:} Current work quantifies resolution-switch asymmetry but should integrate multifaceted factors like MTP latency, stutter frequency, spatial distortion, and multi-sensory interactions like audio-visual synchronization \cite{anwar2020measuring,fei2019qoe}. Future models must also account for individual variability, analyzing how demographics, VR expertise, or content type (e.g., fast-paced gaming vs. static educational tours) modulate sensitivity to degradations. Cross-modal sensing (e.g., combining physiological signals with gaze tracking) could also refine QoE models. Longitudinal studies tracking neurophysiological habituation to quality fluctuations could further refine dynamic weighting strategies.

\textbf{Deployment Architecture Integration:} Seamless integration of intelligent transmission schemes into next-generation wireless architectures warrants exploration. Deploying the proposed DRL framework at scale requires harmonization with evolving standards like OpenRAN \cite{agarwal2022qoe}. Here, splitting intelligence between centralized units (for long-term QoE optimization) and distributed RAN controllers (for real-time resource adaptation) could balance responsiveness with computational efficiency. Cross-layer orchestration via open interfaces would enable interoperability across heterogeneous networks while supporting massive concurrent VR sessions. Challenges include reconciling centralized DRL training with distributed execution across disaggregated RAN units while maintaining millisecond-scale responsiveness.

\textbf{Security and Privacy:} The user-centric paradigm necessitates deep visibility into user states (e.g., physiological responses, movement patterns), raising critical privacy concerns. Future work must embed privacy-by-design principles \cite{yeznabad2024qoe}. Federated learning could train QoE models on-device without raw data exposure (e.g., lowering discomfort scores instead of raw EEG), while homomorphic encryption might enable secure QoE metric computation. Regulatory-compliant frameworks for handling biometric data, especially in cross-border edge clouds, are essential to strike a balance between personalisation and ethical constraints.

\bibliographystyle{IEEEtran}
\bibliography{ref_mag}

\begin{thebibliography}{10}
\providecommand{\url}[1]{#1}
\csname url@samestyle\endcsname
\providecommand{\newblock}{\relax}
\providecommand{\bibinfo}[2]{#2}
\providecommand{\BIBentrySTDinterwordspacing}{\spaceskip=0pt\relax}
\providecommand{\BIBentryALTinterwordstretchfactor}{4}
\providecommand{\BIBentryALTinterwordspacing}{\spaceskip=\fontdimen2\font plus
\BIBentryALTinterwordstretchfactor\fontdimen3\font minus
  \fontdimen4\font\relax}
\providecommand{\BIBforeignlanguage}[2]{{%
\expandafter\ifx\csname l@#1\endcsname\relax
\typeout{** WARNING: IEEEtran.bst: No hyphenation pattern has been}%
\typeout{** loaded for the language `#1'. Using the pattern for}%
\typeout{** the default language instead.}%
\else
\language=\csname l@#1\endcsname
\fi
#2}}
\providecommand{\BIBdecl}{\relax}
\BIBdecl

\bibitem{yaqoob2020survey}
A.~Yaqoob, T.~Bi, and G.-M. Muntean, ``{A survey on adaptive 360 video
  streaming: Solutions, challenges and opportunities},'' \emph{IEEE Commun.
  Surv. Tutor.}, vol.~22, no.~4, pp. 2801--2838, 2020.

\bibitem{10907861}
Y.~Chen, H.~Lu, L.~Qin, C.~Wu, and C.~W. Chen, ``Streaming 360° {VR} video
  with statistical {QoS} provisioning in mmwave networks from delay and rate
  perspectives,'' \emph{IEEE Trans. Wireless Commun.}, vol.~24, no.~6, pp.
  4721--4737, 2025.

\bibitem{anwar2020measuring}
M.~S. Anwar, J.~Wang, A.~Ullah, W.~Khan, S.~Ahmad, and Z.~Fei, ``{Measuring
  quality of experience for 360-degree videos in virtual reality},'' \emph{Sci.
  China Inf. Sci}, vol.~63, no.~10, p. 202301, 2020.

\bibitem{fei2019qoe}
Z.~Fei, F.~Wang, J.~Wang, and X.~Xie, ``{QoE evaluation methods for 360-degree
  VR video transmission},'' \emph{IEEE J. Sel. Top. Sign. Proces}, vol.~14,
  no.~1, pp. 78--88, 2019.

\bibitem{zuo2022adaptive}
X.~Zuo, J.~Yang, M.~Wang, and Y.~Cui, ``{Adaptive bitrate with user-level QoE
  preference for video streaming},'' in \emph{IEEE INFOCOM 2022-IEEE Conference
  on Computer Communications}.\hskip 1em plus 0.5em minus 0.4em\relax IEEE,
  2022, pp. 1279--1288.

\bibitem{johansson2014self}
I.~Johansson, ``{Self-clocked rate adaptation for conversational video in
  LTE},'' in \emph{Proceedings of the 2014 ACM SIGCOMM workshop on Capacity
  sharing workshop}, 2014, pp. 51--56.

\bibitem{maura2024experimenting}
F.~Maura, M.~Casasnovas, and B.~Bellalta, ``{Experimenting with adaptive
  bitrate algorithms for virtual reality streaming over Wi-Fi},'' in
  \emph{Proceedings of the 30th Annual International Conference on Mobile
  Computing and Networking}, 2024, pp. 1930--1937.

\bibitem{bampis2021towards}
C.~G. Bampis, Z.~Li, I.~Katsavounidis, T.-Y. Huang, C.~Ekanadham, and A.~C.
  Bovik, ``Towards perceptually optimized adaptive video streaming-a realistic
  quality of experience database,'' \emph{IEEE Trans. Image Process.}, vol.~30,
  pp. 5182--5197, 2021.

\bibitem{agarwal2022qoe}
B.~Agarwal, M.~A. Togou, M.~Ruffini, and G.-M. Muntean, ``{QoE-driven
  optimization in 5g o-ran-enabled hetnets for enhanced video service
  quality},'' \emph{IEEE Commun. Mag}, vol.~61, no.~1, pp. 56--62, 2022.

\bibitem{3gpp_technical_38214}
3GPP, ``{NR; Physical layer procedures for data;},'' {3rd Generation
  Partnership Project (3GPP)}, Technical Specification (TS) 38.214, Jul 2025,
  version 19.0.0.

\bibitem{bentaleb2018survey}
A.~Bentaleb, B.~Taani, A.~C. Begen, C.~Timmerer, and R.~Zimmermann, ``{A survey
  on bitrate adaptation schemes for streaming media over HTTP},'' \emph{IEEE
  Commun. Surv. Tutor.}, vol.~21, no.~1, pp. 562--585, 2018.

\bibitem{yeznabad2024qoe}
Y.~F. Yeznabad, M.~Helfert, and G.-M. Muntean, ``{QoE-driven cross-layer
  bitrate allocation approach for MEC-supported adaptive video streaming},''
  \emph{IEEE Trans. Netw. Serv. Manage.}, 2024.

\bibitem{vidaurre2009time}
C.~Vidaurre, N.~Kr{\"a}mer, B.~Blankertz, and A.~Schl{\"o}gl, ``{Time domain
  parameters as a feature for EEG-based brain--computer interfaces},''
  \emph{Neural Netw.}, vol.~22, no.~9, pp. 1313--1319, 2009.

\bibitem{blankertz2011single}
B.~Blankertz, S.~Lemm, M.~Treder, S.~Haufe, and K.-R. M{\"u}ller,
  ``{Single-trial analysis and classification of ERP components—a
  tutorial},'' \emph{NeuroImage}, vol.~56, no.~2, pp. 814--825, 2011.

\bibitem{schulman2017proximal}
J.~Schulman, F.~Wolski, P.~Dhariwal, A.~Radford, and O.~Klimov, ``{Proximal
  policy optimization algorithms},'' \emph{arXiv preprint arXiv:1707.06347},
  2017.

\end{thebibliography}

\section{Biography}

\begin{IEEEbiographynophoto}{Chang Wu}
(Graduate Student Member) is currently pursuing the Ph.D. degree in the Department of EEIS, USTC. His research interests include UCN, B5G/6G, intelligent RAN, traffic scheduling, and congestion control.
\end{IEEEbiographynophoto}

\begin{IEEEbiographynophoto}{Yuang Chen} is currently pursuing the Ph.D. degree in the Department of EEIS, USTC. His research interests include VR streaming and xURLLC.
\end{IEEEbiographynophoto}

\begin{IEEEbiographynophoto}{Yiyuan Chen}
is currently pursuing the Ph.D. degree in the Department of EEIS, USTC. His research interests mainly focus on AI-based biological signal processing.
\end{IEEEbiographynophoto}

\begin{IEEEbiographynophoto}{Fengqian Guo} is currently an Associate Researcher with the School of Information Science and Technology at USTC. His research interests include wireless resource optimization and multimedia transmission.
\end{IEEEbiographynophoto}

\begin{IEEEbiographynophoto}{Xiaowei Qin} has been an Associate Professor with the Key Laboratory of Wireless-Optical Communications of the Chinese Academy of Sciences, USTC, since 2014. His research interests include big data in mobile communication networks and terminal low energy consumption.
\end{IEEEbiographynophoto}

\begin{IEEEbiographynophoto}{Hancheng Lu}
 (Senior Member, IEEE) is currently a full Professor with the Department of EEIS, USTC. His research interests include resource optimization in wireless communication systems and caching and service offloading at wireless network edges.
\end{IEEEbiographynophoto}

\vfill

\end{document}